\documentclass{aa}
\usepackage{graphicx}
\begin{document}
   \title{[C II] emission and star formation in late-type galaxies. II\\
          A model
   \thanks{Based on observations with the Infrared Space Observatory (ISO),
          an ESA project with instruments funded by ESA member states
          (especially the PI countries: France, Germany, the Netherlands
          and the United Kingdom) and with the participation of ISAS
          and NASA.}}


   \author{D. Pierini\inst{1,2}
          \and
          K.J. Leech\inst{3}
          \and
          H.J. V\"olk\inst{4}
          }

   \offprints{D. Pierini}

   \institute{Ritter Astrophysical Research Center, The University of Toledo,
              Toledo, OH 43606
         \and Max-Planck-Institut f\"ur extraterrestrische Physik,
              Giessenbachstrasse, D-85748 Garching\\
              \email{dpierini@mpe.mpg.de}
         \and ISO Data Centre, Astrophysics Division, ESA Space Science Dept.,
              P.O. Box 50727 Madrid\\
              \email{kleech@iso.vilspa.esa.es}
         \and Max-Planck-Institut f\"ur Kernphysik,
              Saupfercheckweg 1, D-69117 Heidelberg\\
              \email{Heinrich.Voelk@mpi-hd.mpg.de}
             }

   \date{Received ..., 2001; accepted ..., 2002}

   \abstract{
    We study the relationship between gas cooling via the [C II]
    ($\rm \lambda = 158~\mu m$) line emission and dust cooling
    via the far-IR continuum emission on the global scale of a galaxy
    in normal (i.e. non-AGN dominated and non-starburst) late-type systems.
    It is known that the luminosity ratio of total gas and dust cooling,
    $L_{\mathrm{C~II}}/L_{\mathrm{FIR}}$, shows a non-linear behaviour with
    the equivalent width of the $\rm H \alpha$ ($\rm \lambda = 6563~\AA$)
    line emission, the ratio decreasing in galaxies of lower massive
    star-formation activity.
    This result holds despite the fact that known individual Galactic
    and extragalactic sources of the [C II] line emission show
    different [C II] line-to-far-IR continuum emission ratios.
    This non-linear behaviour is reproduced by a simple quantitative model
    of gas and dust heating from different stellar populations, assuming that
    the photoelectric effect on dust, induced by far-UV photons,
    is the dominant mechanism of gas heating in the general diffuse
    interstellar medium of the galaxies under investigation.
    According to the model, the global $L_{\mathrm{C~II}}/L_{\mathrm{FIR}}$
    provides a direct measure of the fractional amount of non-ionizing UV
    light in the interstellar radiation field and not of the efficiency
    of the photoelectric heating.
    The model also defines a method to constrain the stellar initial mass
    function from measurements of $L_{\mathrm{C~II}}$ and $L_{\mathrm{FIR}}$.
    A sample of 20 Virgo cluster galaxies observed in the [C II] line
    with the Long Wavelength Spectrometer on board the Infrared Space
    Observatory is used to illustrate the model.
    The limited statistics and the necessary assumptions behind
    the determination of the global [C II] luminosities from the spatially
    limited data do not allow us to establish definitive conclusions
    but data-sets available in the future will allow tests of both
    the reliability of the assumptions of our model and the statistical
    significance of our results.
   }

   \maketitle
%

\section{Introduction}
   The [C II]($\rm ^2 P_{3/2}~-~^2 P_{1/2}$) ($\rm \lambda = 157.7409~\mu m$)
   transition of singly ionized carbon is globally the dominant cooling line
   in the general diffuse Interstellar Medium (ISM), excluding H II regions
   (Dalgarno \& McCray 1972; Tielens \& Hollenbach 1985;
   Wolfire et al. 1995).
   Excitation of the $\rm ^2 P_{3/2}$ level of $\rm C^{+}$ is due
   to inelastic collisions with either neutral hydrogen atoms,
   molecules or electrons (Dalgarno \& McCray 1972; Stacey 1985;
   Kulkarni \& Heiles 1987).

   Heating of the neutral interstellar gas is thought to be mainly due
   to photoelectrons (de Jong 1980) emitted by dust grains
   and Polycyclic Aromatic Hydrocarbons that are exposed to ultraviolet
   radiation from stars, both in the general diffuse ISM (Heiles 1994;
   Wolfire et al. 1995) and in the denser standard Photodissociation Regions
   (PDRs), at the interfaces between molecular clouds and H II regions
   (Tielens \& Hollenbach 1985; Bakes \& Tielens 1998).
   This photoelectric effect is essentially due to non-ionizing
   far-UV photons with $\rm h \nu \ge 6~eV$, as determined by
   the typical values of the work functions of the grain surfaces.
   In the field of a galaxy, this radiation is dominated by B3 to B0 stars
   with $\rm 5 \leq M \leq 20~M_{\mathrm{\sun}}$ (e.g. Xu et al. 1994).
   In addition, hotter, more massive stars will contribute locally.

   Earlier studies of the innermost regions of gas-rich
   and starburst galaxies (Crawford et al. 1985; Stacey et al. 1991;
   Carral et al. 1994) have found that the [C II] line intensity,
   $I_{\mathrm{C~II}}$, is typically a few $\rm \times~10^{-3}$
   of the far-IR continuum emission from the dust,
   heated by the stellar continuum emission.
   This fraction is of the same order of magnitude
   as the theoretical efficiency of the photoelectric heating,
   suggesting that the ratio between the far-IR continuum emission
   from dust and the stellar far-UV emission is constant
   (e.g. Kaufman et al. 1999).

   With the advent of the Long Wavelength Spectrometer (LWS)
   (Clegg et al. 1996) on-board the ESA's Infrared Space Observatory (ISO)
   (Kessler et al. 1996), it has been possible to detect
   the [C II] line emission of large samples of normal,
   i.e. non AGN-dominated and non-starburst, galaxies
   (e.g. Malhotra et al. 1997, 2000, 2001; Smith \& Madden 1997;
   Leech et al. 1999).
   Based on the last two data-sets and on that of Stacey et al. (1991),
   Pierini et al. (1999 -- hereafter referred to as P99) have discovered
   a non-linear dependence of the ratio between the [C II] line intensity
   and the total far-IR continuum intensity FIR (as defined in Helou
   et al. 1985), $I_{\mathrm{C~II}}/I_{\mathrm{FIR}}$, on the equivalent width
   of the $\rm H \alpha$ line emission at $\rm \lambda = 6562.8~\AA$
   ($\rm H \alpha~EW$).
   The latter is proportional to the ratio between the stellar
   Lyman continuum and the stellar red continuum and is a tracer
   of the recent mass-normalized star formation rate (SFR),
   where mass normalization is in terms of the stellar red luminosity,
   and is linked to the star formation history
   of an individual galaxy (Kennicutt et al. 1994).
   In particular, P99 found that $I_{\mathrm{C~II}}/I_{\mathrm{FIR}}$ is about
   $\rm 4 \times 10^{-3}$ in the ``normal star-forming'' galaxies
   (with $\rm H \alpha~EW \ge 10~\AA$), while it decreases continuously
   with decreasing $\rm H \alpha~EW$ for ``quiescent'' galaxies
   (with $\rm H \alpha~EW < 10~\AA$).
   They qualitatively interpreted this drop in terms of a dominant
   ``[C II]-quiet'' component of the far-IR emission in quiescent
   late-type galaxies, which are mainly identified
   with early-type spirals (cf. Leech et al. 1999).
   In these galaxies, dust heating is increasingly dominated by the general
   interstellar radiation field (ISRF), due to stars of low mass
   ($\rm M < 5~M_{\mathrm{\sun}}$) (e.g. Xu et al. 1994).
   However, stars with $\rm M < 5~M_{\mathrm{\sun}}$ can not produce
   any substantial photoelectric effect on dust grains, for physical reasons
   which have to do with the depth of the Fermi level in the solids
   that constitute dust grains.
   In this sense such galaxies are ``[C II]-quiet''.
   The scenario of P99 has been recently invoked by Malhotra et al. (2000)
   in order to explain the low values of
   $L_{\mathrm{C~II}}/L_{\mathrm{FIR}}$ (1.2 -- $\rm 2.2 \times 10^{-3}$)
   of four nearby E/S0 galaxies observed with LWS.

   Boselli et al. (2002) discuss the use of the [C II] luminosity
   as a diagnostic of the SFR in non IR-luminous galaxies, in place of
   e.g. the $\rm H \alpha$ luminosity, on the basis of
   {\it empirical} correlations.
   Their investigation carries on the seminal study of Stacey et al. (1991)
   and the later one of P99.
   They also review the empirical knowledge of the relationship between
   the [C II] line emission and the far-IR continuum emission from dust
   in different interstellar media and compare the [C II] line-to-far-IR
   continuum emission ratio for different Galactic and extragalactic sources
   of the [C II] line emission in more details than in P99.

   In this paper, we present and discuss a simple but nontrivial
   {\it theoretical model} for the global energetics (per unit of
   galactic mass) of gas cooling via the [C II] line emission
   and dust cooling via far-IR emission in normal galaxies.
   This model is aimed at interpreting the relationship between
   global [C II] line-to-far-IR continuum emission ratio
   and massive star formation activity per unit of mass for normal galaxies,
   as suggested by the results of P99 and Malhotra et al. (2000).
   In particular, the trend found by P99 holds despite the fact that
   individual sources of the [C II] line emission have different
   [C II] line-to-far-IR continuum emission ratios (Stacey et al. 1991; P99;
   Boselli et al. 2002).
   The model presented here is due to H.J. V\"olk.
   It employs two moments of the stellar initial mass function (IMF)
   and two corresponding averages of the star formation rate (SFR)
   per unit of mass.
   We adopt standard characterizations of the IMF, assumed to be universal,
   in agreement with much of the evidence available in the literature
   (e.g. Meyer et al. 2000 and references therein) and with the expectations
   of self-regulating star formation models of disk-galaxies
   (e.g. Silk 1997).
   By its very nature, this model does not account for the dependence
   of the [C II] line emission on the astrophysical properties
   of the individual sources of this emission, like density of the gas
   and far-UV intensity of the local radiation field.
   For this kind of investigations we refer the reader to Heiles (1994),
   Wolfire et al. (1995) Kaufman et al. (1999) and references therein.

   The two samples of Virgo cluster member late-type galaxies observed
   by Leech et al. (1999) and by Smith \& Madden (1997) provide us with
   a total sample of 24 normal spirals with a large dynamic range
   in SF history and with homogeneous measurements of the observables
   of our interest (cf. Sect. 2).
   These two samples are weakly affected by Malmquist bias,
   because of the cluster depth (e.g. Gavazzi et al. 1999).

   The previous LWS measurements are used to constrain the model,
   after a correction for the limited aperture of LWS (not introduced by P99),
   described and discussed in Sect. 2.
   This correction is admittedly not fully certain but is physically
   well motivated.
   In addition, the corrected [C II] line-to-far-IR continuum emission ratios
   are consistent with those determined within the LWS beam area, directly from
   the previous LWS data and the ISOPHOT (Lemke et al. 1996) data of
   Tuffs et al. (2002), for 10 of the sample galaxies (cf. Sect. 2).
   Finally, the correction leads to a result consistent with that of P99,
   as shown in Sect. 3.

   The limited statistics of the data available to us nevertheless
   do not allow us to definitively confirm or disprove the model
   and its quantitative consequences for the IMF.
   However, the model has an intrinsic astrophysical value and larger data
   sets, which will be available in the near future, will allow
   a detailed test.
   Therefore, the reader interested exclusively in the model may skip
   Sect. 2 and 3 and go directly to Sect. 4, where we introduce the model,
   and from there to Sect. 5, where we interpret the behaviour of
   $L_{\mathrm{C~II}}/L_{\mathrm{FIR}}$ with the mass-normalized massive SFR,
   reproduced in Sect. 3 (see also P99).
   A discussion of our results is contained in Sect. 6, while Sect. 7
   gives a summary of our conclusions.
%

\section{The galaxy sample and complementary data}

\subsection{The Sample}
   The ISO Guaranteed Time program VIRGO combines studies of a deep
   optically complete, volume-limited sample of spiral, irregular
   and blue compact dwarf galaxies, selected from the Virgo Cluster Catalogue
   (VCC) of Binggeli et al. (1985) (Tuffs et al. 2002).
   In particular, 19 spiral galaxies
   (18 with $\rm B_{\mathrm{T}} \le 12.3~mag$, plus NGC\,4491 with
   $\rm B_{\mathrm{T}} = 13.4~mag$) could be observed by Leech et al.
   (1999 -- hereafter referred to as L99) with LWS in the spectral region
   around the [C II] line.
   Fourteen galaxies were detected in the [C II] line at a signal-to-noise
   ratio above $\rm 3 \sigma$ and upper limits were set to the other 5.
   Assuming a distance of 21~Mpc from Virgo, the 70\arcsec~LWS half power
   beam-width (HPBW) corresponds to $\sim$ 7~kpc and encompasses at least
   one exponential disk scale-length for most of the sample galaxies,
   with the exception of those with an optical major axis larger than 5 arcmin.
   We note that a slightly larger (80\arcsec) beam size has been determined
   by Lloyd (2000) for LWS.
   However, the discrepancy in the adopted LWS beam size does not affect
   the following results.

   Five further VCC spiral galaxies with
   $\rm 12.5 \le B_{\mathrm{T}} \le 13.9~mag$ were observed by
   Smith \& Madden (1997 -- hereafter referred to as SM97)
   with LWS in the [C II] line region.
   The latter 5 galaxies complement the previous 19, so that
   the total sample of 24 Virgo cluster member galaxies represents
   normal spirals with a large dynamic range in mass-normalized massive SFR
   ($\rm H \alpha~EW \le 71~\AA$) and in morphological type
   (from S0a to Sd).

   According to P99, we divide the total sample of galaxies into quiescent
   and normal star-forming galaxies according to the values
   of $\rm H \alpha~EW < 10~\AA$ and $\rm \ge 10~\AA$, respectively.
   $\rm H \alpha +$[N II](6548, 6583 $\rm \AA$) equivalent widths
   are available for 20 out of 24 (20/24) galaxies either from long-slit
   (from 3 to 7 arcmin) spectroscopy of their central region
   (Kennicutt \& Kent 1983) or from CCD imaging
   (G. Gavazzi, private communication).

   All the 19 galaxies of the L99 subsample have been imaged
   by Boselli et al. (1997) in the $\rm K^{\prime}$-band
   ($\rm \lambda = 2.1~\mu m$) and in the H-band ($\rm \lambda = 1.65~\mu m$).
   Since the 5 galaxies of the SM97 subsample lack CCD measurements
   in these two near-IR bands, we convert their VCC total B-band
   ($\rm \lambda = 0.55~\mu m$) magnitudes into $\rm K^{\prime}$-band ones
   via the relation between the colour index $\rm B_T-K^{\prime}$
   and the Hubble type found by Boselli et al. (1997).
   Furthermore, we convert these $\rm K^{\prime}$-band magnitudes
   into H-band ones via the average $\rm H-K^{\prime}$ colour
   of spiral galaxies of all Hubble types (0.26 mag), found by these authors.

   18/24 galaxies have total nonthermal radio fluxes at 1.4 GHz, listed in
   Gavazzi \& Boselli (1999).
   For 21/24 objects, we have selected detections/upper limits
   for their total IRAS 60 $\rm \mu m$ and 100 $\rm \mu m$ fluxes from
   either Rice et al. (1988) or Soifer et al. (1989) or Moshir et al. (1990),
   according to their relative optical sizes with respect to the IRAS HPBW.

   The parameters of the total sample of 24 VCC late-type galaxies
   observed with LWS, relevant to this analysis, are given in Tab. 1
   as follows: \newline
   Column 1: the NGC and VCC denominations; \newline
   Column 2: the morphological type, as listed in the VCC; \newline
   Column 3: logarithm of the length of major axis, as listed in
   de Vaucouleurs et al. (1991 -- hereafter referred to as RC3);
   \newline
   Column 4: logarithm of the axial ratio, as listed in the RC3; \newline
   Column 5: the $\rm H \alpha +$[N II] equivalent width
   ($\rm H \alpha~EW$ for short); \newline
   Column 6: the total H-band magnitude, determined and corrected
   for Galactic extintion and galaxy inclination according to
   Gavazzi \& Boselli (1996); \newline
   Column 7: the total nonthermal radio flux at 1.4 GHz; \newline
   Column 8: the total IRAS flux at 60 $\rm \mu m$; \newline
   Column 9: the total IRAS flux at 100 $\rm \mu m$; \newline
   Column 10: the observed [C II] line flux; \newline
   Column 11: references of the [C II] data.
   \begin{table*}
      \caption[]{Galaxy parameters}
         \label{Tab1}
     $$ 
         \begin{array}{p{0.15\linewidth}cccrrrrrrc}
            \hline
            \noalign{\smallskip}
            Denomination & {\mathrm{Hubble~type}}~~ & log~D & log~R & H{\mathrm{\alpha}}~EW & H~~~ & F_{1.4 GHz} & F_{60 {\mathrm{\mu}} m}~ & F_{100 {\mathrm{\mu}} m} & F_{C~II}~~~~~~~ & {\mathrm{ref}} \\
            NGC~/~VCC & & ~~0.1^{\prime}~ & & {[\mathrm{\AA}]}~~~~ & {[\mathrm{mag}]}~ & {[\mathrm{mJy}]}~~ & {[\mathrm{Jy}]}~~~ & {[\mathrm{Jy}]}~~~ & 10^{-20}~{[\mathrm{W~cm^{-2}}]} & \\
            \noalign{\smallskip}
            \hline
            \noalign{\smallskip}
NGC\,4178~~VCC\,66 & {\mathrm{SBc}} & 1.71 & 0.45 & 23~~~~~ &  8.89~ &  26.2~~~ &  2.11~~ &  8.08~~ &  7.80~~~~~~~~ & {\mathrm{L99}} \\
NGC\,4189~~VCC\,89 & {\mathrm{SBc}} & 1.38 & 0.14 & 20~~~~~ &  9.35~ &  17.1~~~ &  3.05~~ &  8.93~~ &  9.10~~~~~~~~ & {\mathrm{SM97}} \\
NGC\,4192~~VCC\,92$^{\mathrm{a}}$ & {\mathrm{Sb:}} & 1.99 & 0.55 &  9~~~~~ &  6.50~ &  73.3~~~ &  8.11~~ & 23.07~~ &  8.80~~~~~~~~ & {\mathrm{L99}} \\
NGC\,4222~~VCC\,187 & {\mathrm{Scd}} & 1.52 & 0.86 &  7~~~~~ & 10.93~ &   3.9~~~ &  0.99~~ &  3.19~~ &  4.50~~~~~~~~ & {\mathrm{SM97}} \\
NGC\,4293~~VCC\,460 & {\mathrm{Sa pec}} & 1.75 & 0.34 &  2~~~~~ &  7.29~ &  19.1~~~ &  4.58~~ & 10.43~~ &  2.70~~~~~~~~ & {\mathrm{L99}} \\
NGC\,4294~~VCC\,465 & {\mathrm{SBc}} & 1.51 & 0.42 & 55~~~~~ &  9.34~ &  22.5~~~ &  2.73~~ & < 5.91~~ &  8.40~~~~~~~~ & {\mathrm{SM97}} \\
NGC\,4299~~VCC\,491 & {\mathrm{Scd}} & 1.24 & 0.03 & 74~~~~~ & 10.12~ &  18.6~~~ &  2.63~~ & < 5.39~~ &  6.80~~~~~~~~ & {\mathrm{SM97}} \\
NGC\,4394~~VCC\,857$^{\mathrm{b}}$ & {\mathrm{SBb}} & 1.56 & 0.05 &  2~~~~~ &  8.26~ &  -~~~ &  0.95~~ &  4.02~~ &  1.40~~~~~~~~ & {\mathrm{L99}} \\
NGC\,4402~~VCC\,873 & {\mathrm{Sc}} & 1.59 & 0.55 & 16~~~~~ &  8.42~ &  59.5~~~ &  5.43~~ & 17.48~~ & 17.60~~~~~~~~ & {\mathrm{L99}} \\
NGC\,4429~~VCC\,1003 & {\mathrm{S0/Sa pec}} & 1.75 & 0.34 &  -~~~~~ &  6.85~ &  -~~~ &  1.54~~ &  4.31~~ &  3.8~~~~~~~~ & {\mathrm{L99}} \\
NGC\,4438~~VCC\,1043$^{\mathrm{c}}$ & {\mathrm{Sb (tides)}} & 1.93 & 0.43 &  6~~~~~ &  7.14~ & 148.9~~~ &  3.76~~ & 11.27~~ &  8.20~~~~~~~~ & {\mathrm{L99}} \\
NGC\,4450~~VCC\,1110$^{\mathrm{d}}$ & {\mathrm{Sab pec}} & 1.72 & 0.13 &  2~~~~~ &  6.97~ &  10.2~~~ &  1.34~~ &  6.95~~ &  2.10~~~~~~~~ & {\mathrm{L99}} \\
NGC\,4461~~VCC\,1158 & {\mathrm{Sa}} & 1.55 & 0.39 &  -~~~~~ &  8.02~ &  -~~~ &  -~~ &  -~~ & < 2.10~~~~~~~~ & {\mathrm{L99}} \\
NGC\,4477~~VCC\,1253 & {\mathrm{SB0/SBa}} & 1.58 & 0.04 &  -~~~~~ &  7.43~ &  -~~~ &  0.54~~ &  1.18~~ & < 1.60~~~~~~~~ & {\mathrm{L99}} \\
NGC\,4491~~VCC\,1326 & {\mathrm{SBa}} & 1.23 & 0.30 &  0~~~~~ &  9.92~ &  -~~~ &  2.77~~ &  3.49~~ &  1.30~~~~~~~~ & {\mathrm{L99}} \\
NGC\,4503~~VCC\,1412 & {\mathrm{Sa}} & 1.55 & 0.33 & 2~~~~~ &  8.05~ &  -~~~ &  -~~ &  -~~ & < 1.30~~~~~~~~ & {\mathrm{L99}} \\
NGC\,4522~~VCC\,1516 & {\mathrm{Sc/Sb}} & 1.57 & 0.57 & 10~~~~~ &  9.42~ &  23.4~~~ &  1.30~~ &  4.20~~ &  8.10~~~~~~~~ & {\mathrm{SM97}} \\
NGC\,4569~~VCC\,1690$^{\mathrm{e}}$ & {\mathrm{Sab}} & 1.98 & 0.34 &  2~~~~~ &  6.67~ &  72.5~~~ & 10.08~~ & 26.60~~ & 15.10~~~~~~~~ & {\mathrm{L99}} \\
NGC\,4579~~VCC\,1727$^{\mathrm{f}}$ & {\mathrm{Sab}} & 1.77 & 0.10 &  4~~~~~ &  6.61~ &  97.4~~~ &  5.85~~ & 20.86~~ &  7.80~~~~~~~~ & {\mathrm{L99}} \\
NGC\,4596~~VCC\,1813 & {\mathrm{SBa}} & 1.60 & 0.13 &  2~~~~~ &  7.30~ &  -~~~ &  0.49~~ &  1.28~~ & < 2.00~~~~~~~~ & {\mathrm{L99}} \\
NGC\,4608~~VCC\,1869 & {\mathrm{SB0/a}} & 1.51 & 0.08 &  -~~~~~ &  8.03~ & < 2.8~~~ &  -~~ &  -~~ & < 0.60~~~~~~~~ & {\mathrm{L99}} \\
NGC\,4647~~VCC\,1972 & {\mathrm{Sc}} & 1.46 & 0.10 & 16~~~~~ &  8.59~ &  56.3~~~ &  5.35~~ & 16.04~~ & 19.00~~~~~~~~ & {\mathrm{L99}} \\
NGC\,4654~~VCC\,1987 & {\mathrm{SBc}} & 1.69 & 0.24 & 30~~~~~ &  7.79~ & 125.3~~~ & 13.93~~ & 37.16~~ & 28.40~~~~~~~~ & {\mathrm{L99}} \\
NGC\,4698~~VCC\,2070$^{\mathrm{g}}$ & {\mathrm{Sa}} & 1.60 & 0.21 &  6~~~~~ &  7.55~ &  -~~~ &  0.26~~ & 1.86~~ &  1.30~~~~~~~~ & {\mathrm{L99}} \\
            \noalign{\smallskip}
            \hline
         \end{array}
     $$ 
\begin{list}{}{}
\item[$^{\mathrm{a}}$] LINER (Rauscher 1995; Barth et al. 1998)
\item[$^{\mathrm{b}}$] LINER (Keel 1983; Rauscher 1995)
\item[$^{\mathrm{c}}$] LINER (Ho et al. 1997)
\item[$^{\mathrm{d}}$] LINER (Gonzales-Delgado et al. 1997)
\item[$^{\mathrm{e}}$] LINER/Sy (Stauffer 1982; Keel 1983; Ho et al. 1997)
\item[$^{\mathrm{f}}$] LINER/Sy 1.9 (Stauffer 1982; Ho et al. 1997)
\item[$^{\mathrm{g}}$] Sy 2 (Ho et al. 1997)
\end{list}
   \end{table*}

   In Tab. 1, notes identify the 7 early-type galaxies which are claimed
   to host nuclear activity typical of a LINER/AGN.
   The presence of an AGN increases both the [C II] line emission
   and the nonthermal radio emission of these galaxies
   in a way that is hard to predict quantitatively.
   As an example, however, an extra nonthermal radio emission due to
   the production of energetic electrons {\it without star formation}
   by an AGN will lead to secondary electrons on grains which will act like
   photoelectrons heating the gas.
   For NGC\,4394 (Keel 1983), NGC\,4569 and NGC\,4579 (Stauffer 1982),
   claims of non-stellar nuclear activity were raised before the selection
   and observation of our sample.
   On the other hand, NGC\,4192 (Barth et al. 1998), NGC\,4569 and NGC\,4579
   (Ho et al. 1997) have recently been defined as
   transition spirals, while NGC\,4438 (Ho et al. 1997)
   seems to be a marginal candidate for non-stellar nuclear activity.
   The phenomenology of the Virgo cluster spiral galaxy nuclear regions
   has not yet been established (e.g. Rauscher 1995), since different
   types of LINER (i.e. photoionized by a stellar continuum or by an active
   galactic nucleus) can not easily be distinguished from one another
   (e.g. Alonso-Herrero et al. 2000).
   Moreover, these 7 galaxies have values of the IRAS far-IR colour
   f(60)/f(100) in the range 0.28 -- 0.58, which indicates dust colour
   temperatures not particularly warm, and consistent both with
   the average far-IR colours of AGNs (0.58, with a dispersion of 0.20)
   and of non-AGN galaxies (0.40, with a dispersion of 0.12) found by
   Bothun et al. (1989).
   Given this, we still consider it reasonable to include these 7 VCC galaxies
   as part of our sample of normal late-type galaxies.

   On the other hand, the 8 VCC Sc/Scd galaxies of our sample, detected both
   in the [C II] line and at 1.4 GHz, are not known to host any LINER/AGN
   activity.
   These galaxies define the subsample of normal star-forming
   galaxies.
%

\subsection{Data analysis}
   The best way to get the [C II] line-to-far-IR continuum emission ratio
   on a galactic scale is to integrate over complete maps in each emission.
   Since this is not possible at the moment of writing, two options are left.
   One would be to take the [C II] pointings, relate the [C II] flux in them
   to the far-IR flux in a similar sized beam and take this ratio
   as representative of the whole galaxy, as suggested by the referee.
   IRAS beams are messy, so this is not the best way, but, for 10 of
   the galaxies listed in Tab. 1 and detected in the [C II] line emission
   (i.e. NGC\,4178, 4192, 4293, 4394, 4429, 4438, 4450, 4491, 4569 and 4579),
   high signal-to-noise strip maps exist at 60 and $\rm 100~\mu m$, obtained by
   Tuffs et al. (2002) with the 40 arcsec HPBW instrument ISOPHOT
   (Lemke et al. 1996).
   Tuffs et al. have extracted the ISOPHOT far-IR photometry by fitting a model
   of the far-IR surface brightness distribution to the data.
   Thus, for these 10 galaxies, [C II] line-to-far-IR continuum
   emission ratios may be measured within the LWS beam size.
   However, it is not straightforward to conclude that these ratios
   represent the true total ratios.
   The second option is to adopt a [C II] line ``growth curve'',
   but such a tool is not available at the moment of writing.
   A coarse spatial distribution of the [C II] line emission is known only
   for few individual galaxies, where it is found to be complex
   (e.g. Madden et al. 1993).
   
   We decided that a way out of this ``empasse'' is to take the IRAS fluxes
   as the total far-IR ones and determine an aperture correction for
   the observed [C II] line fluxes on the basis of the most commonly accepted
   physical interpretation of the [C II] line emission.
   We will test the total [C II] line-to-far-IR continuum emission ratios
   obtained with our method from the LWS and IRAS measurements against
   those obtained with the method suggested by the referee, i.e.
   from the LWS and ISOPHOT data in our possession.

   We adopt a simple analytical formula to extrapolate the total [C II] line
   flux from the observed one, which is based on the assumptions that
   i) the average ratio between the total [C II] line luminosity and the total
   1.4 GHz radio continuum luminosity, as results of massive star formation
   activity, is constant in galaxies of the same Hubble type and that
   ii) the radial surface brightness profile of the [C II] line emission
   is homologous in galaxies of different Hubble type.
   We note that at this relatively low frequency, the contribution of
   the thermal Bremsstrahlung to the radio continuum emission of
   late-type galaxies is weak (e.g. Gioia et al. 1982).

   In the normal star-forming galaxies, almost the same young stellar
   populations responsible for the gas heating (Sect. 1) dominate
   the SN rate (cf. X94), and, therefore, the production of
   Cosmic Ray electrons and nonthermal radio continuum emission.
   Since the corresponding UV radiation is essentially absorbed inside
   the galaxy by the dust grains, also the total number of photoelectrons
   and therefore the gas heating per $\rm C^+$-atom is {\it only} dependent
   on the UV luminosity and thus on the formation rate of massive stars.
   Almost the same is true for the total nonthermal radio emission
   (Lisenfeld et al. 1996).
   This carries over to the quiescent galaxies, although the fraction
   of SN Ia to SN II plus SN Ib may be larger than in the previous objects
   (cf. X94).
   These galaxies are close to becoming or are in fact already ``radio-quiet''
   (Condon et al. 1991) as well as [C II]-quiet (P99).
   Finally, whatever the massive star formation activity of the galaxy is,
   gas cooling and SNe are disk phenomena in predominantly non-interacting
   galaxies like ours (cf. the case of the ``Antennae'' studied by
   Nikola et al. 1998).
   These astrophysical considerations support our assumptions.

   It may sound dangerous to extrapolate the total [C II] line flux
   from the radio synchrotron emission in order to interpret the total
   gas-to-dust cooling flux ratio, given the existence of the radio/far-IR
   correlation (Rickard \& Harvey 1984; Dickey \& Salpeter 1984;
   de Jong et al. 1985; Helou et al. 1985; see also Condon et al. 1991; X94).
   However, the aperture correction is necessary and we consider
   this extrapolation as a very reasonably physically motivated way
   to achieve it.
   This correction has a larger impact for the larger galaxies of our sample,
   which are expected to have both larger radio and far-IR continuum
   emissions, as a consequence of well-established scaling effects (e.g. X94).
   Most of the galaxies largely undersampled by LWS are early-types,
   as shown by their values of the ``coverage factor'' $CF$ (defined as
   the ratio between the area of the LWS beam and the projected optical
   galaxy area, as defined in the RC3) between 10 and 100 per cent
   (Pierini et al. 2001).
   Nonetheless, for these objects, the behaviour of the total
   $L_{\mathrm{C~II}}/L_{\mathrm{FIR}}$ with the $\rm H \alpha~EW$
   (Sect. 3) is consistent with that reported by P99, where no aperture
   correction to the LWS data was applied.
   On the basis of this consistency, we conclude that the aperture correction
   here introduced does not bias our following conclusions.
   \begin{figure}
   \centering
   \includegraphics[width=9cm]{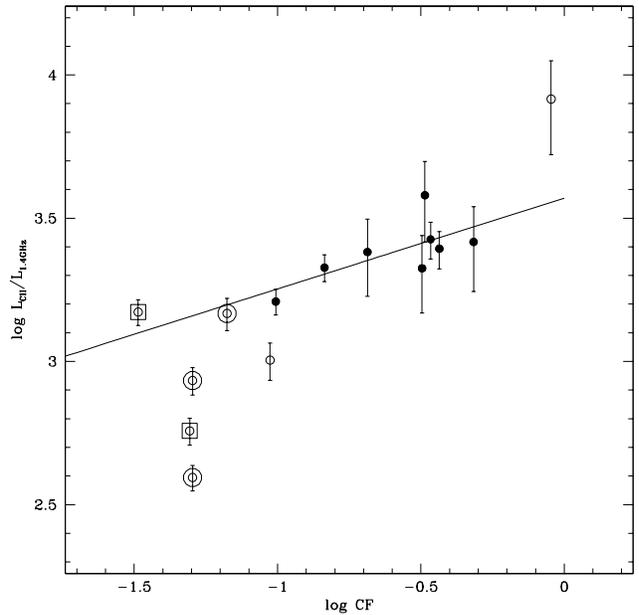}
      \caption{The ratio between the [C II] line luminosity, as derived from
               the {\it observed} flux, and the total radio continuum
               luminosity at 1.4 GHz,
               $L_{\mathrm{C~II}}/L_{\mathrm{1.4\,GHz}}$,
               vs. the ``coverage factor'' $CF$, as defined in the text.
               Hereafter, filled circles represent the normal star-forming
               galaxies, empty circles represent the quiescent ones
               and asterisks represent galaxies without available
               $\rm H \alpha~EW$ (see text).
               Large empty circles and squares mark objects identified
               as Seyfert/LINER galaxies, respectively, while arrows mark
               lower and upper limits.
               Only the 15 galaxies with detected fluxes in both
               the emissions are plotted.
               The solid line represents Equ. 1.
              }
         \label{FigVibStab}
   \end{figure}

   The LWS beamsize probed the [C II] line emission of the 8 VCC Sc/Scd
   galaxies of our sample up to relatively large galactocentric distances
   (Pierini et al. 2001).
   Moreover, there are no existing claims of LINER/AGN activity
   for these galaxies.
   For all these reasons, the subsample defined by these 8 Sc/Scd galaxies
   is adopted to investigate the dependence of the ratio between
   the [C II] line emission (within the LWS beam) and the total radio
   continuum luminosity, $L_{\mathrm{C~II}}/L_{\mathrm{1.4\,GHz}}$,
   on the $CF$.

   As a functional form for the LWS aperture correction, we adopt
   the simple linear equation $y = a~x + b$,
   where $y = log~L_{\mathrm{C~II}}/L_{\mathrm{1.4\,GHz}}$ and $x = log~CF$.
   According to our assumptions,
   $log~L_{\mathrm{C~II}}/L_{\mathrm{1.4\,GHz}}$ is expected
   to decrease with decreasing values of $log~CF$ as a consequence only
   of the reduced fraction of the [C II] line emission probed by LWS,
   so that the parameter $a$ is independent of the Hubble type.
   In contrast, the constant term $b$, defining the intrinsic ratio between
   the total [C II] line luminosity and the total 1.4 GHz radio continuum
   luminosity, depends to first order on the average star formation history
   of a galaxy, i.e. on the Hubble type of the galaxy (Kennicutt et al. 1994).
   We may also expect that $b$ depends to second order on the presence
   of an AGN and on the relationship between global properties of the ISM
   and Hubble type, if any.

   Fig. 1 shows the plot of $log~L_{\mathrm{C~II}}/L_{\mathrm{1.4\,GHz}}$
   vs. $log~CF$ for all the galaxies with available obervables,
   $L_{\mathrm{C\,II}}$ being determined from the observed [C II] line flux
   assuming a distance of 21 Mpc for Virgo.
   Hereafter, quiescent galaxies, normal star-forming galaxies
   and galaxies without measured $\rm H \alpha~EW$ are represented by
   empty circles, filled circles and asterisks, respectively.
   We also mark those galaxies with claimed non-stellar nuclear activity
   (cf. Sect. 2.1) by large circles and squares, respectively.
   In Fig. 1, the solid line reproduces the least-squares fit of equation
    \begin{equation}
       log~\frac{L_{\mathrm{C~II}}}{L_{\mathrm{1.4\,GHz}}} = 0.32 (\pm 0.21)~log~CF + 3.57 (\pm 0.13)\,,
    \end{equation}
   obtained for the 8 Sc/Scd galaxies.

   We note that the parameter $a$ is different from 0 only at
   the $\rm 1 \sigma$ level, maybe as a result of the limited statistics.
   Nevertheless, we correct the observed [C II] line fluxes of all the galaxies
   (listed in Col. 10 of Tab. 1) to total ones via the multiplicative term
   $10^{0.32~\times~log~CF}$.
   \begin{figure}
   \centering
   \includegraphics[width=9cm]{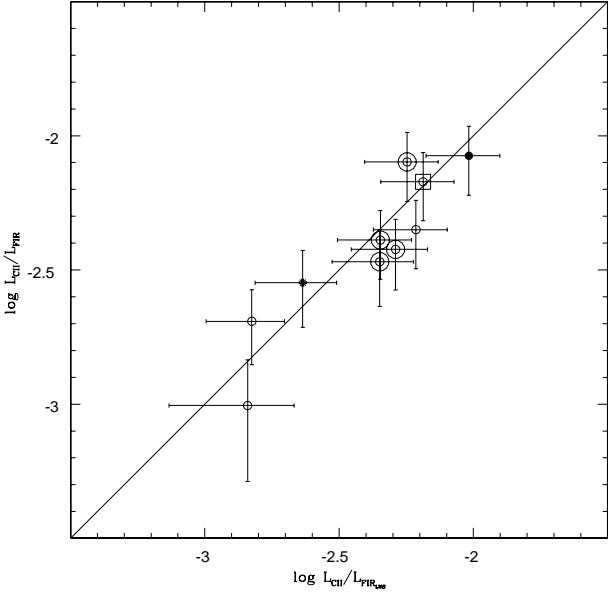}
      \caption{Comparison of the ratio of the total [C II] line luminosity,
               obtained from the correction for the LWS aperture (Equ. 1)
               and from the total IRAS far-IR continuum luminosity FIR
               (as defined by Helou et al. 1985),
               $L_{\mathrm{C~II}}/L_{\mathrm{FIR}}$, and the ratio of
               the observed [C II] line luminosity and the ISOPHOT far-IR
               continuum luminosity FIR within the LWS beam area,
               $L_{\mathrm{C~II}}/L_{\mathrm{FIR_{\mathrm{LWS}}}}$.
               Only the 10 galaxies with available LWS, ISOPHOT and IRAS
               measurements are plotted (see text).
              }
         \label{Fig2}
   \end{figure}

   The method previously illustrated is admittedly crude but its feasibility
   may be tested against the data,  as previously mentioned.
   In fact, ratios between the observed [C II] line luminosity and the far-IR
   continuum luminosity FIR (Helou et al. 1985) within the LWS beam
   may be obtained from the existing ISOPHOT photometry (Tuffs et al. 2002).
   Tuffs et al. show that the ISOPHOT photometry at 60 and $\rm 100~\mu m$
   is consistent with the analogous IRAS photometry.   
   Thus we have obtained aperture photometry of 10 sample galaxies
   with available LWS, ISOPHOT and IRAS data (i.e. NGC\,4178, 4192, 4293, 4394,
   4429, 4438, 4450, 4491, 4569 and 4579) at 60 and $\rm 100~\mu m$,
   after deconvolving the models of Tuffs et al. with the ISOPHOT beam,
   assumed to be reproduced by a two-dimensional circular gaussian
   of 40 arcsec full-width-at-half-maximum (FWHM).
   
   Fig. 2 shows the comparison between the {\it total} [C II] line-to-far-IR
   FIR emission ratios ($log~L_{\mathrm{C~II}}/L_{\mathrm{FIR}}$ on the y-axis)
   and the [C II] line-to-far-IR FIR emission ratios within the LWS beam area
   ($log~L_{\mathrm{C~II}}/L_{\mathrm{FIR_{\mathrm{LWS}}}}$ on the x-axis).
   The total ratio $log~L_{\mathrm{C~II}}/L_{\mathrm{FIR}}$ is obtained
   by applying the aperture correction (Equ. 1) to the observed LWS [C II] line
   flux and from the total IRAS fluxes at 60 and $\rm 100~\mu m$, according
   to Helou et al. (1985).
   Conversely, $log~L_{\mathrm{C~II}}/L_{\mathrm{FIR_{\mathrm{LWS}}}}$
   is obtained from the observed [C II] line flux and from the integration
   of the far-IR surface brightness distribution at 60 and $\rm 100~\mu m$,
   as obtained with ISOPHOT, within the LWS beam area.
   We assume an error of 5 per cent in the model ISOPHOT fluxes
   and an error of 30 per cent in the ISOPHOT absolute calibration.

   The ratios estimated through these two different methods are consistent
   within the uncertainties and the approximations in both methods,
   especially when considering the difference between a total value
   and a rather local one.
   This is not trivial.
   Thus, we conclude that most of the emission both in the [C II] line
   and in the far-IR continuum comes from the regions probed by LWS,
   at least for our sample galaxies, as assumed by P99.
   ``A posteriori'' this result supports the assumptions behind
   the correction for the LWS aperture, especially when considering that
   9 of the previous 10 galaxies are not among those used to determine
   this aperture correction.
   Thus, we conclude that the estimates of the total [C II] line fluxes,
   obtained through the aperture correction of Equ. 1, are robust.
   \begin{figure*}
   \centering
   \includegraphics[width=12cm]{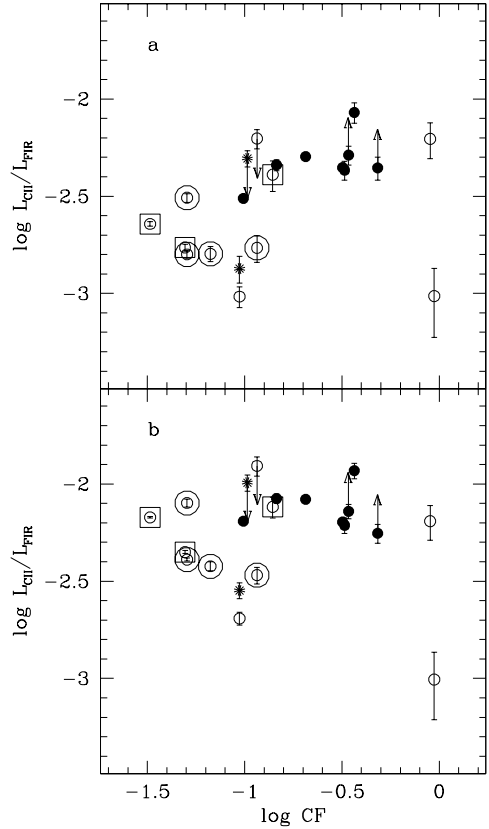}
      \caption{The ratio between the [C II] line luminosity
               and the total far-IR continuum luminosity FIR (as defined by
               Helou et al. 1985), $L_{\mathrm{C~II}}/L_{\mathrm{FIR}}$,
               vs. $CF$, {\it prior to} ({\bf a}) and {\it after}
               ({\bf b}) the correction for the LWS aperture.
               Only the 21 galaxies with available measurements are plotted.
              }
         \label{Fig3}
   \end{figure*}
%

\section{$L_{\mathrm{C~II}}/L_{\mathrm{FIR}}$ vs. $\rm H \alpha~EW$ in normal
late-type galaxies}
   In Fig. 3a,b, we plot the ratio between the [C II] line
   luminosity and the total far-IR continuum luminosity FIR
   (Helou et al. 1985), $log~L_{\mathrm{C~II}}/L_{\mathrm{FIR}}$,
   vs. $log~CF$ prior to and after the correction for the LWS aperture
   effects, respectively.
   Although this correction affects the values of
   $L_{\mathrm{C\,II}}/L_{\mathrm{FIR}}$ of individual galaxies,
   there is no definite global trend of $L_{\mathrm{C~II}}/L_{\mathrm{FIR}}$
   with $CF$ in Fig. 3a and, a fortiori, in Fig. 3b.

   The early-type spiral galaxies with claimed non-stellar nuclear activity
   (cf. Sect. 2.1) are among those with the lowest values
   of $L_{\mathrm{C~II}}/L_{\mathrm{FIR}}$ in Fig. 3a,b,
   but their total $L_{\mathrm{C~II}}/L_{\mathrm{FIR}}$ (Fig. 3b)
   is on average more than one order of magnitude higher than that of
   the ``[C II]-deficient'' galaxies identified by Malhotra et al. (1997),
   with IRAS far-IR colours between 0.6 and 1.4.
   For the same galaxies, the mean value of
   $L_{\mathrm{C~II}}/L_{\mathrm{FIR}}$ is intermediate between the range
   1.2 -- $\rm 2.2 \times 10^{-3}$ spanned by 4 nearby E/S0 galaxies
   observed with LWS (Malhotra et al. 2000) and the mean value of
   $L_{\mathrm{C~II}}/L_{\mathrm{FIR}}$ of VCC galaxies
   of later Hubble types (Fig. 3b).
   This result suggests that the non-stellar nuclear activity claimed for
   some of our galaxies does not affect systematically their values
   of $L_{\mathrm{C~II}}/L_{\mathrm{FIR}}$.
   We conclude that $L_{\mathrm{C~II}}/L_{\mathrm{FIR}}$ is, on average,
   lower for quiescent galaxies than for normal star-forming ones,
   even though the scatter is large.

  The average ratio between the [C II] line emission and the far-IR
  continuum emission is consistent with the theoretical efficiency
  of the photoelectric heating of the gas, as found by previous authors
  (cf. Sect. 1).

  Fig. 4 shows the dependence of the total
  $L_{\mathrm{C~II}}/L_{\mathrm{FIR}}$ on the $\rm H \alpha~EW$ for
  the 19/24 galaxies of our sample with available observables.
  Here, the galaxy out of bounds is NGC\,4491,
  with $\rm H \alpha~EW$ consistent with $\rm 0~\AA$.

  The distribution in Fig. 4 is much more regular than the one
  reproduced in Fig. 3b.
  For the 6 normal star-forming galaxies detected both in the [C II] line
  and in the far-IR, the mean value of
  $L_{\mathrm{C~II}}/L_{\mathrm{FIR}}$ is equal to $\rm 7.9 \times 10^{-3}$
  with a dispersion of $2.1 \times 10^{-3}$.
  This figure is still consistent with the analogous mean value
  of $\sim 4 \times 10^{-3}$ obtained by P99 for an heterogeneous sample
  of normal star-forming galaxies without any kind of aperture correction.
  For the quiescent galaxies, $L_{\mathrm{C~II}}/L_{\mathrm{FIR}}$ drops
  continuously with decreasing values of the $\rm H \alpha~EW$,
  in agreement with P99.
  Finally, galaxies with claimed non-stellar nuclear activity do not behave
  in a peculiar way in Fig. 4.

  These results strengthen those obtained by P99.
  In fact, P99 assumed that the bulk of the IRAS far-IR continuum
  emission of their VCC galaxies originated within the region
  sampled by the aperture of LWS (set equal to 80~arcsec) for each galaxy.
  P99 assumed implicitly that the characteristic FWHM of the IRAS
  $\rm 60~\mu m$ and $\rm 100~\mu m$ continuum emissions
  of normal galaxies was defined by the IRAS observations of the nearby
  Sab galaxy NGC\,5713 (see Lord et al. 1996), which indicate that
  the FWHM of the IRAS $\rm 25~\mu m$ and $\rm 60~\mu m$ emissions
  of this galaxy is about 1/5 of its optical diameter.
  The LWS HPBW indeed encompasses 1/5 of the galaxy's optical diameter
  in VCC galaxies with $\rm log~D \le 1.76$, i.e. for all the objects
  observed with LWS, with the exception of NGC\,4192, NGC\,4438
  and NGC\,4569 (cf. Tab. 1).
  As seen in Sect. 2.2 this assumption seems to hold for the present case.
  Conversely, the values of $L_{\mathrm{C~II}}/L_{\mathrm{FIR}}$ adopted by
  P99 for their subsample of galaxies observed by Stacey et al. (1991)
  did not suffer because of a marked difference in beam size
  of the [C II] and far-IR observations.
  Here, we confirm that the dependence of
  $L_{\mathrm{C~II}}/L_{\mathrm{FIR}}$ on the $\rm H \alpha~EW$ found by P99
  is not due either to their assumption or to the inclusion of LINER/AGN
  candidates in their sample.
   \begin{figure}
   \centering
   \includegraphics[width=9cm]{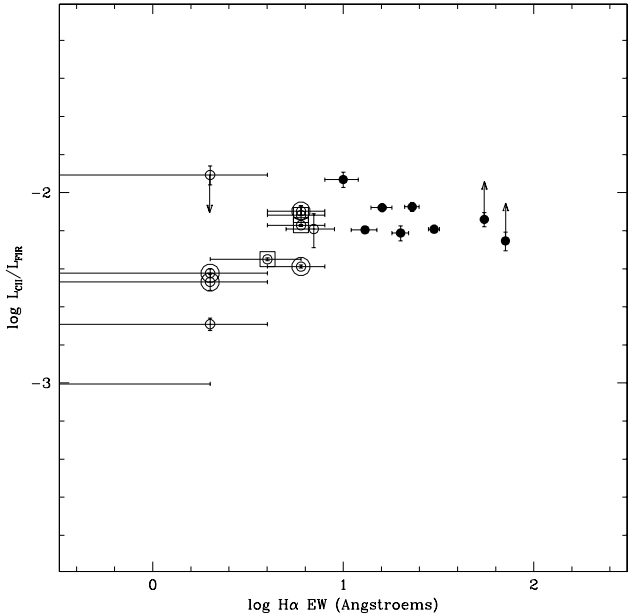}
      \caption{$L_{\mathrm{C~II}}/L_{\mathrm{FIR}}$ vs. the equivalent width
               of the $\rm H \alpha$ line, $\rm H \alpha~EW$.
               $L_{\mathrm{C~II}}$ is corrected for the LWS aperture.
               Only the 19 galaxies with available measurements are plotted.
               The galaxy out of bounds is NGC\,4491, with $\rm H \alpha~EW$
               consistent with $\rm 0~\AA$.
               The non-linear relation between
               $L_{\mathrm{C~II}}/L_{\mathrm{FIR}}$ and $\rm H \alpha~EW$
               found by P99 is confirmed.
              }
         \label{Fig4}
   \end{figure}
%

\section{The model}
   P99 interpreted the non-linear relation between
   $L_{\mathrm{C~II}}/L_{\mathrm{FIR}}$ and the $\rm H \alpha~EW$
   under the assumption that photoelectric heating is the dominant mechanism
   of gas heating in the diffuse ISM of normal spiral galaxies,
   on the scale of a galaxy.
   The existence of a lower characteristic far-UV photon energy threshold
   for the photoelectric effect on the dust, coupled to the different
   massive SFRs of galaxies of different morphology
   (cf. Kennicutt et al. 1994), qualitatively explains why
   quiescent early-type spiral galaxies have lower values
   of $L_{\mathrm{C~II}}/L_{\mathrm{FIR}}$ than normal star-forming
   late-type ones.

   Here, we introduce a simple quantitative model in order to describe
   this behaviour, based both on generally accepted concepts
   of galaxy star formation history and on observations.
   We adopt mass-normalized quantities, in order to eliminate
   simple scaling effects among galaxies of different masses.
   Mass-normalization is in terms of the total near-IR luminosity.
   The latter is, in fact, a good estimate of the bulk of the mass
   of the old stellar population, which dominates the mass
   of the ``luminous matter'' of giant galaxies (Aaronson et al. 1979).

   In our modeling, we adopt the formalism introduced by Xu et al.
   (1994 -- hereafter referred to as X94), in their interpretation
   of the properties of the radio/far-IR correlation
   for normal late-type galaxies.
   We refer the reader to X94 for further details and discussion of
   the physical justification of the parameterization introduced hereafter.

   The galaxy mass $M$ is estimated as follows:
   \begin{equation}
      M = \frac{L_{\mathrm{H}}}{2.05 \times 10^{25}~{\mathrm{[W\,\mu m^{-1}]}}}~~~~(M_{\mathrm{{\sun}}})\,,
   \end{equation}
   where $L_{\mathrm{H}}$ is the H-band luminosity.

   The mass-normalized total IRAS far-IR continuum (FIR) emission within
   the wavelength range of 40 -- 120 $\rm \mu m$ is expressed through
   the following formula:
   \begin{equation}
      \frac{L_{\mathrm{FIR}}}{M} = \zeta~\int_{m_1}^{m_2} \phi (m)~P (m)~L (m)~\tau (m)~\overline{s (\tau (m))}~dm\,,
   \end{equation}
   with $\zeta = 1/1.8$ denoting the fraction of the FIR emission
   contained within 40 -- 120 $\rm \mu m$, which is quite a constant factor
   among galaxies (Helou et al. 1985), $m_1 = \rm 1~M_{\mathrm{\sun}}$ and
   $m_2 = \rm 100~M_{\mathrm{\sun}}$ specifying the mass range of stars
   which are responsible for the heating of the dust (we ignore stars
   less massive than the Sun because they contribute little to
   the dust heating and take an upper cut-off of the IMF
   of $\rm 100~M_{\mathrm{\sun}}$).
   The stellar initial mass function (IMF) is denoted by $\phi (m)$,
   the probability of the light of a star of mass $m$ to be absorbed by dust
   by $P (m)$.
   Finally, $L (m)$ and $\tau (m)$ are the Main Sequence (MS)
   luminosity and life time of a star of mass $m$, respectively, and
   \begin{equation}
      \overline{s (\tau (m))} = \frac{ \int_{0}^{\tau (m)} s (t)~dt}{\tau (m)}
   \end{equation}
   is the star formation rate (per unit of mass) averaged over the life time
   of this star.

   The total energy emitted during the MS life of a star of given mass $m$,
   $L (m) \times \tau (m)$, is given by:
   \begin{equation}
      L (m)~\tau (m) = \left\{ \begin{array}{ll}
                                 10^{9.95}~m~\mathrm{[L_{\mathrm{\sun}}\,yr]}      & \mbox{$1 \le m < \rm 5~M_{\mathrm{\sun}}$} \\
                                 10^{9.6}~m^{3/2}~\mathrm{[L_{\mathrm{\sun}}\,yr]} & \mbox{$5 \le m < \rm 100~M_{\mathrm{\sun}}$.}
                               \end{array}
                       \right .
   \end{equation}
   According to X94, 60\% of the radiation from stars more massive than
   $\rm 5~M_{\mathrm{\sun}}$ is absorbed and then reradiated in the far-IR
   by the dust, while the absorption probability of radiation from stars
   with $1 \le m < \rm 5~M_{\mathrm{\sun}}$ is 30\%.

   We consider six different IMFs (Salpeter 1955; Miller \& Scalo 1979;
   Kennicutt 1983; Kroupa et al. 1993; X94; Scalo 1998) and, in each case,
   we assume that the IMF is universal (cf. Meyer et al. 2000)
   and independent of the SFR.
   For each IMF, we calculate the normalization factor from the definition:
   \begin{equation}
      \int_{m_{\mathrm{min}}}^{m_{\mathrm{max}}} \phi (m)~m~dm~=~1\,,
   \end{equation}
   imposing continuity at the values of $m$ corresponding to a change
   of the IMF slope and taking the lower stellar mass
   $m_{\mathrm{min}} = \rm 1~M_{\mathrm{\sun}}$ and the upper stellar mass
   $m_{\mathrm{max}} = \rm 100~M_{\mathrm{\sun}}$.

   Thus we rewrite $L_{\mathrm{FIR}}/M$ as:
   \begin{equation}
      \frac{L_{\mathrm{FIR}}}{M} = \frac{L_{\mathrm{FIR}}}{M}_{\mathrm{loud}}+\frac{L_{\mathrm{FIR}}}{M}_{\mathrm{quiet}}= a~s_8 + b~s_9\,,
   \end{equation}
   where the suffix ``loud'' (``quiet'') defines the far-IR component due to
   stars more (less) massive than $\rm 5~M_{\mathrm{\sun}}$,
   which are (are not) responsible for the photoelectric heating (Sect. 1).

   The quantities $s_8$ and $s_9$ (in units of $\rm yr^{-1}$) define
   the star formation rates (per unit of mass) averaged over times
   of $\rm 10^8~yrs$ and $\rm 3 \times 10^9~yrs$, respectively.
   MS life times shorter than $\rm 10^8~yrs$ correspond to stars
   more massive than $\rm 5~M_{\mathrm{\sun}}$, while MS life times between
   $\rm 10^8~yrs$ and few $\rm \times 10^9~yrs$ characterize stars
   with $1 \le m < \rm 5~M_{\mathrm{\sun}}$.
   Given the previous assumptions, the quantities $a$ and $b$ (in units of
   $\rm L_{\mathrm{\sun}}~{M_{\mathrm{\sun}}}^{-1}~yr$) depend only on the IMF.
   We discuss this interesting aspect in further detail in Sect. 5.2.
   In Tab. 2, we show the values of the parameters $a$ and $b$,
   obtained from Equ. 3 through 6, corresponding to the adopted IMFs.
   \begin{table}
      \caption[]{``a'' and ``b'' as functions of the IMF}
         \label{Tab2}
     $$ 
         \begin{array}{p{0.3\linewidth}ccc}
            \hline
            \noalign{\smallskip}
            IMF & a & b & \frac{a}{a+b}\\
              & {[\mathrm{10^8~L_{\sun} {M_{\sun}}^{-1} yr}]} & {[\mathrm{10^8~L_{\sun} {M_{\sun}}^{-1} yr}]} & \\
            \noalign{\smallskip}
            \hline
            \noalign{\smallskip}
            Salpeter '55 & 3.85 & 1.10 & 0.78\\
            Miller \& Scalo '79 & 5.21 & 3.70 & 0.58\\
            Kennicutt '83 & 8.94 & 3.69 & 0.71\\
            Kroupa et al. '93 & 3.37 & 2.23 & 0.60\\
            Xu et al. '94 & 9.26 & 3.34 & 0.73\\
            Scalo '98 & 8.73 & 4.08 & 0.68\\
            \noalign{\smallskip}
            \hline
         \end{array}
     $$
   \end{table}

   The mass-normalized total [C II] line luminosity
   is expressed as follows:
   \begin{equation}
      \frac{L_{\mathrm{C~II}}}{M} = c~s_8\,.
   \end{equation}
   The parameter $c$ (in units of
   $\rm L_{\mathrm{\sun}}~{M_{\mathrm{\sun}}}^{-1}~yr$) is assumed to depend
   on the IMF and, to second order, on the efficiency of the photoelectric
   effect in galaxies of different far-UV field strengths
   (cf. Bakes \& Tielens 1994), i.e. of different morphological
   classification.
   Eventually, the latter dependence concerns effects due to metallicity
   and properties of dust and ISM, whether these properties are dependent
   on Hubble type.
   This formula provides a physical justification for the use of
   the [C II] line emission as a tracer of star formation
   in non AGN-dominated and non-starburst galaxies.
   This aspect is discussed by Boselli et al. (2002) and not here.

   Finally, from Equ. 7 and 8 it follows that:
   \begin{equation}
      \frac{L_{\mathrm{C~II}}}{L_{\mathrm{FIR}}} = \frac{c}{a} \times \frac{s}{s+\frac{b}{a}}\,,
   \end{equation}
   where
   \begin{equation}
      s = \frac{s_8}{s_9}\,.
   \end{equation}
   According to these two equations, $L_{\mathrm{C~II}}/L_{\mathrm{FIR}}$
   tends to $c/a$ for $s >> b/a$ and to $(c/b)~s$ for $s << b/a$.
%

\section{Results}

\subsection{On the relation between gas cooling, dust cooling
and star formation history of individual galaxies}
   In our model, the ratio between gas cooling via the [C II] line emission
   and dust cooling via the far-IR continuum emission depends only on
   the parameter $c$ and on the two average SFRs (per unit of mass)
   $s_8$ and $s_9$, once the IMF is chosen.
   Hereafter, we adopt the X94 IMF for illustrative purposes.

   We assume that galaxies of the same Hubble type (HT)
   have the same characteristic SF time scale.
   Under the further assumption that $s_9$ is the {\it median}
   of the values of $s_8$ of individual galaxies of the same morphology
   (X94), we derive the {\it median} values of $s_9$ and $c$
   for each Hubble type as follows:
   \begin{equation}
      s_9 = \rm \frac{<\frac{L_{\mathrm{FIR}}}{M}>_{\mathrm{HT}}}{a+b}\,,
   \end{equation}
   \begin{equation}
      c = \frac{<\frac{L_{\mathrm{C~II}}}{M}>_{\mathrm{HT}}}{s_9}\,.
   \end{equation}
   We remind the reader that the median is especially appropriate
   as a measure of central tendency of a skewed distribution of data.
   For the present illustrative purposes, we estimate the {\it mean}
   values of $L_{FIR}/M$ and $L_{C~II}/M$ for any given morphological type
   represented by the galaxy sample described in Sect. 2.1 instead.
   This choice allows the comparison of the mean value of $L_{C~II}/L_{FIR}$
   found for the normal star-forming galaxies to that found by P99
   for similar objects.

   In Fig. 5, we plot the mass-normalized total [C II] luminosity
   vs. the mass-normalized total far-IR continuum luminosity
   for the 20 Virgo cluster galaxies with available suitable observables.
   In Fig. 5a,b,c only the 18 objects in the three groups Sa/Sab, Sb/Sbc,
   Sc/Scd are shown, since only for these three groups of morphological types
   we can constrain our model.
   \begin{figure*}
   \centering
   \includegraphics[width=12cm]{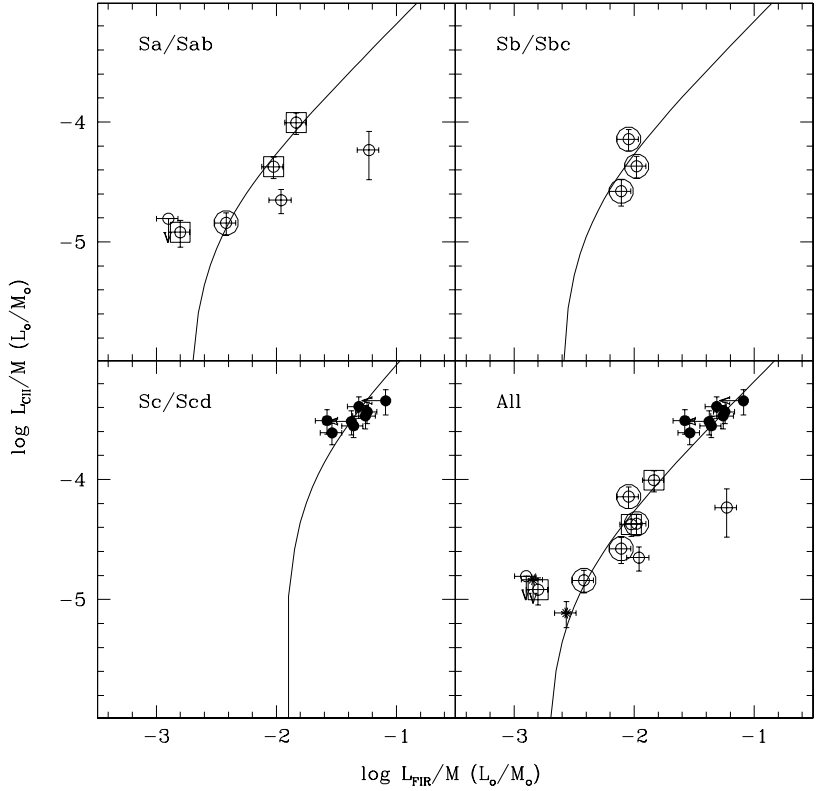}
      \caption{The mass-normalized total (i.e. corrected for the LWS aperture)
               [C II] luminosity, $L_{\mathrm{C~II}}/M$, vs.
               the mass-normalized total far-IR continuum luminosity FIR,
               $L_{\mathrm{FIR}}/M$.
               The 18 galaxies of the three groups of morphological classes
               for which the available observables may constrain the model,
               i.e. Sa/Sab ({\bf a}), Sb/Sbc ({\bf b}) and Sc/Scd ({\bf c}),
               are plotted.
               In each panel, the solid line represents the values
               of $L_{\mathrm{C~II}}/M$ obtained from Equ. 13.
               All the 20 galaxies with available suitable observables
               are plotted, superposed to the relation expressed by Equ. 13
               for the Sa/Sab group ({\bf d}).
               It is evident that the available statistics are not sufficent
               to fully characterize the dependence of the model parameters
               on the galaxy morphology.
              }
         \label{Fig5}
   \end{figure*}
   In Fig. 5d we show all the 20 objects superimposed on the model curve
   obtained for the Sa/Sab galaxies as a reference.
   In each panel, the solid line represents the logarithmic value
   of $L_{C~II}/M$ obtained as follows:
   \begin{equation}
      log~\frac{L_{\mathrm{C~II}}}{M} = log~( \frac{L_{\mathrm{FIR}}}{M} - b~s_9) + log~\frac{c}{a}\,.
   \end{equation}

   From inspection of Fig. 5, we obtain that:
   \begin{enumerate}
     \item the model relation between $log~L_{\mathrm{C~II}}/M$
   and $log~L_{\mathrm{FIR}}/M$ is non-linear for any given morphological
   classification;
     \item in the model, $L_{\mathrm{C~II}}/L_{\mathrm{FIR}}$ drops
   with decreasing values of $L_{\mathrm{FIR}}/M$, independent of
   galaxy classification, but this drop takes place at a characteristic value
   of $L_{\mathrm{FIR}}/M \sim b~s_9$, increasing with galaxy lateness;
     \item $L_{\mathrm{C~II}}/L_{\mathrm{FIR}}$ tends to first order to
   the same asymptotic value $c/a$ at sufficiently high values
   of $L_{\mathrm{FIR}}/M$ for galaxies of different Hubble types;
     \item the galaxy distribution shown in Fig. 5 is consistent with
   the relation between current star formation activity and galaxy lateness
   (cf. Kennicutt et al. 1994; Boissier \& Prantzos 2000), even though
   the non-linearity of the model is not well covered by the available
   data points;
     \item the statistics are not yet sufficient to characterize
   the dependence of the model parameters on the galaxy morphology;
     \item therefore the statistics are not yet sufficient to derive
   any quantitative difference induced by the assumed IMF (cf. Sect. 5.2).
   \end{enumerate}

   We emphasize that only large statistical samples can prove or disprove
   the validity of the model.
   Nevertheless these illustrative results are quite encouraging.
%

\subsection{A new method of constraining the stellar IMF}
   The model described in Sect. 4 allows us to express
   the mass-normalized total [C II] and far-IR continuum luminosities
   in terms of combinations of two star formation rates (per unit of mass),
   $s_8$ and $s_9$, averaged over $10^8$ and $3 \times 10^9$ years,
   respectively, weighted by the two parameters $a$ and $b$, depending
   {\it only} on the IMF.

   Galaxies of a given morphological type distribute along a non-linear
   relation in the $log~L_{\mathrm{C~II}}/M$--$log~L_{\mathrm{FIR}}/M$ plane,
   according to the ratio between $s_8$ and $s_9$ in such a way that
   \begin{equation}
      \frac{L_{\mathrm{C~II}}}{L_{\mathrm{FIR}}} = \frac{c}{a+b}
   \end{equation}
   when $s_8 \sim s_9$, and that, conversely,
   \begin{equation}
      \frac{L_{\mathrm{C~II}}}{L_{\mathrm{FIR}}} = \frac{c}{a}
   \end{equation}
   when $s_8$ is large compared to $s_9$.

   If the luminosities involved in these two equations were available
   for a large statistically complete sample of normal galaxies,
   it would be straightforward to derive an accurate value
   of $a/(a+b)$ from Equ. 14 and 15.
   Since $a/(a+b)$ is sensitive to the IMF (cf. Col. 4 of Tab. 2),
   one can effectively constrain the stellar IMF via measurements
   of gas and dust cooling.
%

\section{Discussion}
   According to different studies (de Jong 1980; Tielens \& Hollenbach 1985;
   Bakes \& Tielens 1994, 1998; Wolfire et al. 1995; Kaufman et al. 1999),
   the heating of the interstellar gas is mainly due to collisions
   with photoelectrons ejected from different dust components,
   when illuminated by far-UV photons.
   Contrary to gas heating, dust heating is due also to the general ISRF
   (e.g. Condon et al. 1991; X94; Popescu et al. 2000),
   although the consensus is not unanimous.

   Early detailed models of gas heating (e.g. Tielens \& Hollenbach 1985)
   considered only the relatively dense atomic interstellar environments
   dominated by the far-UV field, where dust heating is also dominated by
   the emission from stars more massive than $\rm 5~M_{\mathrm{\sun}}$.
   Therefore, these models naturally interpreted the ratio between
   gas cooling via the [C II] line emission and dust cooling
   via the far-IR continuum emission as a direct measure of
   the efficiency of the photoelectric effect on dust grains.
   In an analogous way, the early studies of the [C II] line emission
   in external galaxies (Crawford et al. 1985; Stacey et al. 1991;
   Carral et al. 1994) targeted the innermost regions of gas-rich
   and starburst galaxies, where both the fractional content
   of molecular gas (e.g. Pierini et al. 2001) and the contribution
   of the OB associations to the ISRF (e.g. Bronfman et al. 2000) are large.
   Since these stellar populations are associated with dense PDRs,
   it is no surprise that the previous models were successful
   in interpreting the phenomenology and the energetics of gas cooling
   in most of the galaxies observed in the [C II] line region
   for the first time.

   Adopting the PDR models of Kaufman et al. (1999),
   Malhotra et al. (2001) have found that a significant fraction
   of the total [C II] line emission of normal galaxies is associated
   with PDRs very close to star formation regions.
   By contrast, Madden et al. (1993) have shown that the diffuse components
   of the ISM, associated with the atomic hydrogen, are responsible for
   the [C II] line emission on scales larger than a galactic nucleus.
   In the inner regions of a galaxy, the contribution of the diffuse
   ionized medium is also non negligible (Heiles 1994; Negishi et al. 2001).
   These results have been supported by P99, on the basis of the observations
   made by L99 and SM97 with LWS, and confirmed by Pierini et al. (2001),
   via a comparison with models of the diffuse ISM of Wolfire et al. (1995).
   We believe that the solution to this dilemma rests in the differences
   not only among the galaxy regions probed by observations
   (Pierini et al. 2001) but also in the massive star formation activity
   of the sample objects.

   As far as orders of magnitude are concerned, even on a galactic scale
   the average ratio between the [C II] line intensity, as measured by
   the LWS, and the IRAS far-IR continuum emission FIR (Helou et al. 1985)
   is consistent with the theoretical efficiency of gas heating via
   the far-UV light induced photoelectric effect on different dust components
   (cf. Lord et al. 1996).
   However, P99 discovered that, to first order, this ratio is a function
   of the massive star formation activity for normal late-type galaxies.
   This result holds despite the fact that individual Galactic
   and extragalactic sources of the [C II] line emission show different values
   of the [C II] line-to-far-IR emission ratio (e.g. Stacey et al. 1991; P99;
   Boselli et al. 2002).
   In Sect. 5.1 we have reproduced the trend observed by P99 with a model,
   under the assumption (X94) that the time evolution of the SFR
   has an exponential law, where the characteristic time scale
   increases with galaxy lateness.
   Our parametric model is supported both by theoretical results
   on galaxy evolution and by observations.
   As a result, we find that the global $L_{\mathrm{C~II}}/L_{\mathrm{FIR}}$
   is a direct measure of the fractional content of far-UV photons
   in the ISRF of individual normal galaxies, whatever the components
   of the ISM (dense PDRs, diffuse atomic/ionized gas), where gas heating
   takes place, are, and not of the efficiency of the photoelectric heating.

   Since massive star formation declines faster with time in spiral galaxies
   of earlier types, on average, these galaxies are expected to have
   low values of $L_{\mathrm{C~II}}/L_{\mathrm{FIR}}$.
   The opposite is true for the late-type spiral galaxies.
   A particularly low production of far-UV photons (per unit of mass)
   naturally explains the non-detection of 4 early-type spirals,
   none of which associated with an AGN, which were observed by L99
   with LWS in the [C II] line region (cf. Tab. 1 in Sect. 2.1).
   This scenario has been recently invoked by Malhotra et al. (2000)
   in order to qualitatively explain the low values of
   $L_{\mathrm{C~II}}/L_{\mathrm{FIR}}$ (1.2 -- 2.2 $\rm \times 10^{-3}$)
   of 4 nearby E/S0 galaxies observed with LWS, in agreement with
   the earlier interpretation of P99.
   According to the model, the efficiency of the photoelectric heating
   (on a galactic scale) may be derived if the absolute value of
   the global far-UV field strength is known.
   The latter is not simply proportional to the far-IR emission from dust
   (Sect. 4), given the relative contributions of non-ionizing UV and
   the general interstellar radiation field to the dust heating.
%

\section{Conclusions}
   We have investigated the relationship between gas cooling via the [C II]
   ($\rm \lambda = 158~\mu m$) line emission and dust cooling
   via the far-IR continuum emission in normal (i.e. non-AGN dominated
   and non-starburst) late-type galaxies.
   It is known that to first order the luminosity ratio between total gas
   and dust cooling, $L_{\mathrm{C~II}}/L_{\mathrm{FIR}}$, shows
   a non-linear behaviour with the equivalent width of the $\rm H \alpha$
   ($\rm \lambda = 6563~\AA$) line emission, the ratio decreasing
   in galaxies of lower global massive star formation activity.
   This trend holds despite the fact that individual Galactic and extragalactic
   sources of the [C II] line emission show different values of
   the [C II] line-to-far-IR emission ratio.

   This non-linear behaviour is reproduced by a simple quantitative model
   of gas and dust heating from different stellar populations,
   under the assumption that the photoelectric effect on dust induced by
   far-UV photons is the dominant mechanism of gas heating
   in the general diffuse interstellar medium of these galaxies.
   This model employs two moments of the stellar initial mass function (IMF)
   and two corresponding averages of the star formation rate (SFR)
   per unit of mass.
   According to the model, $L_{\mathrm{C~II}}/L_{\mathrm{FIR}}$ directly
   measures the fractional amount of the non-ionizing UV light
   in the interstellar radiation field and not the efficiency
   of the photoelectric heating.

   A sample of 20 Virgo cluster galaxies is used to illustrate the model.
   The insufficient statistics and the assumptions behind the determination
   of the global [C II] luminosities from the spatially limited LWS data
   do not yet allow us to definitively confirm or disprove the model.
   When measurements of the total [C II] line emission and of the other
   quantities invoked by our model will be available for large statistical
   samples of non-AGN and non-starburst galaxies, we shall be able
   to characterize the behaviour of $L_{\mathrm{C~II}}/L_{\mathrm{FIR}}$
   with the $\rm H \alpha~EW$ for each Hubble type and to reproduce it.
   The same wealth of data will allow us to test the astrophysical assumptions
   behind our model, via the comparison of the average star formation rates
   as derived from the [C II] line and other line/continuum emissions.
   Finally, it will also quantitatively constrain the IMF, if universal,
   given the dependence of the results on the IMF.
%

\begin{acknowledgements}
   We are grateful for the support of this work to the
   \emph{Deutsche Agentur f\"ur Raumfahrt Angelegenheiten},
   through \emph{DARA\/} project number 50--OR--9501B.
   D. P. acknowledges support also through Grant NAG5-9202 from
   the National Aeronautics and Space Administration to the University
   of Toledo.
\newline
   We are indebted to G. Gavazzi for providing us with optical data
   prior their publication.
\newline
   We are grateful to the referee, G. J. Stacey, for his useful comments
   which improved the quality of this paper in its final version. 
\end{acknowledgements}


\begin{thebibliography}{}

   \bibitem[1979]{aaronson} Aaronson, M., Mould, J., \& Huchra, J. 1979,
      ApJ, 229, 1

   \bibitem[2000]{alonso} Alonso-Herrero, A., Rieke, M. J.,
      \& Rieke, G. H. 2000,
      ApJ, 530, 688

   \bibitem[1994]{bakes1} Bakes, E. L. O., \& Tielens, A. G. G. M. 1994,
      ApJ, 427, 822

   \bibitem[1998]{bakes2} Bakes, E. L. O., \& Tielens, A. G. G. M. 1998,
      ApJ, 499, 258

   \bibitem[1998]{barth} Barth, A. J., Ho, L. C., Filippenko, A. V.,
      \& Sargent, W. L. W. 1998,
      ApJ, 496, 133

   \bibitem[1985]{binggeli} Binggeli, B., Sandage, A., \& Tammann, G. A. 1985,
      AJ, 90, 1681 (VCC)

   \bibitem[2000]{boissier} Boissier, S., Prantzos, N. 2000,
      MNRAS, 312, 398

   \bibitem[2002]{boselli3} Boselli, A., Gavazzi, G., Lequeux, J.,
      Pierini, D. 2002,
      A\&A, 385, 454

   \bibitem[1998]{boselli2} Boselli, A., Lequeux, J., Sauvage, M.,
      et al. 1998,
      A\&A, 335, 53

   \bibitem[1997]{boselli1} Boselli, A., Tuffs, R. J., Gavazzi, G.,
      et al. 1997,
      A\&AS, 121, 507

   \bibitem[1989]{bothun} Bothun, G. D., Lonsdale, C. J., \& Rice, W. 1989,
      ApJ, 341, 129

   \bibitem[2000]{bronfman} Bronfman, L., Casassus, S., May, J.,
      \& Nyman, L.-$\rm \AA$ 2000,
      A\&A, 358, 521

   \bibitem[1994]{carral} Carral, P., Hollenbach, D. J., Lord, S. D.,
      et al. 1994,
      ApJ, 423, 223

   \bibitem[1996]{clegg} Clegg, P. E., Ade, P. A. R., Armand, C., et al. 1996,
      A\&A, 315, L38

   \bibitem[1991]{condon} Condon, J. J., Anderson, M. L., \& Helou, G. 1991,
      ApJ, 376, 95

   \bibitem[1985]{crawford} Crawford, M. K., Genzel, R., Townes, C. H.,
      \& Watson, D. M. 1985,
      ApJ, 291, 755

   \bibitem[1972]{dalgarno} Dalgarno, A., \& McCray, R. 1972,
      ARA\&A, 10, 375

   \bibitem[1980]{dejong1} de Jong, T. 1980,
      Highlights in Astr., 5, 301

   \bibitem[1985]{dejong2} de Jong, T., Klein, U., Wielebinski, R.,
      \& Wunderlich, E. 1985,
      A\&A, 147, L6

   \bibitem[1991]{devaucouleurs} de Vaucouleurs, G., de Vaucouleurs, A.,
      Corwin, H. G., et al. 1991,
      Third Reference Catalogue of Bright Galaxies,
      (Springer-Verlag, New York) (RC3)

   \bibitem[1984]{dickey} Dickey, J. M., \& Salpeter, E. E. 1984,
      A\&A, 180, 12

   \bibitem[1996]{gavazzi1} Gavazzi, G., \& Boselli, A. 1996,
   in: A UBVJHK photometric catalogue of 1022 galaxies in 8 nearby clusters,
   (Gordon and Breach Science Publishers, New York)

   \bibitem[1999]{gavazzi2} Gavazzi, G., \& Boselli, A. 1999,
      A\&A, 343, 86

   \bibitem[1999]{gavazzi3} Gavazzi, G., Boselli, A., Scodeggio, M.,
      et al. 1999,
      MNRAS, 304, 595

   \bibitem[1982]{gioia} Gioia, I. M., Gregorini, L., \& Klein, U. 1982,
      A\&A, 116, 164

   \bibitem[1997]{gonzales} Gonzales-Delgado, R. M., Perez, E.,
      Tadhunter, C., et al. 1997,
      ApJS, 108, 155

   \bibitem[1994]{heiles} Heiles, C. 1994,
      ApJ, 436, 720

   \bibitem[1985]{helou} Helou, G., Soifer, B. T., \& Rowan-Robinson, M. 1985,
      ApJ, 298, L7

   \bibitem[1997]{ho} Ho, L. C., Filippenko, A. V., \& Sargent, W. L. W. 1997,
      ApJS, 112, 315 

   \bibitem[1999]{kaufman} Kaufman, M. J., Wolfire, M. G., Hollenbach, D. J.,
      \& Luhman M. L. 1999,
      ApJ, 527, 795

   \bibitem[1983]{keel} Keel, W. C. 1983,
      ApJS, 52, 229

   \bibitem[1983]{kennicutt1} Kennicutt, R. C. 1983,
      ApJ, 272, 54

   \bibitem[1983]{kennicutt2} Kennicutt, R. C., \& Kent, S. M. 1983,
      AJ, 88, 1094

   \bibitem[1994]{kennicutt3} Kennicutt, R. C., Tamblyn, P.,
      \& Congdon, C. E. 1994,
      ApJ, 435, 22

   \bibitem[1996]{kessler} Kessler, M. F., Steinz, J. A., Anderegg, M. E.,
      et al. 1996,
      A\&A, 315, L27

   \bibitem[1993]{kroupa} Kroupa, P., Tout, C. A., \& Gilmore, G., 1993,
      MNRAS, 262, 545

   \bibitem[1987]{kulkarni} Kulkarni, S., \& Heiles, C. 1987,
      in: Interstellar Processes,
      eds.\ D. J. Hollenbach \& H. A. Thronson, (Reidel, Dordrecht)

   \bibitem[1999]{leech} Leech, K. J., V\"olk, H. J., Heinrichsen, I.,
      et al. 1999,
      MNRAS, 310, 317 (L99)

   \bibitem[1996]{lemke} Lemke, D., Klaas, U., Abolins, J., et al. 1996,
      A\&A, 315, L64

   \bibitem[1996]{lisenfeld} Lisenfeld, U., V\"olk, H. J., Xu, C. 1996,
      A\&A, 314, 745

  \bibitem[2000]{lloyd} Lloyd, C. 2000,
      in: ISO Beyond Point Sources: Studies of Extended Infrared Emission,
      eds.\ R. J. Laureijs, K. J. Leech \& M. F. Kessler, ESA-SP 455

   \bibitem[1996]{lord} Lord, S. D., Malhotra, S., Lim, T., et al. 1996,
      A\&A, 315, L117

   \bibitem[1993]{madden} Madden, S. C., Geis, N., Genzel, R., et al. 1993,
      ApJ, 407, 579

   \bibitem[1997]{malhotra1} Malhotra, S., Helou, G., Stacey, G., et al. 1997,
      ApJ, 491, L27

   \bibitem[2000]{malhotra2} Malhotra, S., Hollenbach, D., Helou, G.,
      et al. 2000,
      ApJ, 543, 634

   \bibitem[2001]{malhotra3} Malhotra, S., Kaufman, M. J., Hollenbach, D.,
      et al. 2001,
      ApJ, 561, 766

   \bibitem[2000]{meyer} Meyer, M. R., Adams, F. C., Hillenbrand, L. A.,
      et al. 2000,
      in: Protostars and Planets IV, eds.\ V. Mannings, A. P. Boss,
      \& S. S. Russell (University of Arizona Press, Tucson)

   \bibitem[1979]{miller} Miller, G. E., \& Scalo, J. M. 1979,
      ApJS, 41, 513

   \bibitem[1990]{moshir} Moshir, M., Kopan, G., Conrow, T., et al. 1990,
      IRAS Faint Source Catalogue, version 2.0

   \bibitem[2001]{negighi} Negishi, T., Onaka, T., Chan, K.-W.,
      \& Roellig, T. L. 2001,
      A\&A, 375, 566
      
   \bibitem[1998]{nikola} Nikola, T., Genzel, R., Herrmann, F., et al. 1998,
      ApJ, 504, 749

   \bibitem[1999]{pierini1} Pierini, D., Leech, K. J., Tuffs, R. J.,
      \& V\"olk, H. J. 1999,
      MNRAS, 303, L29 (P99)

   \bibitem[2001]{pierini2} Pierini, D., Lequeux, J., Boselli, A.,
      et al. 2001,
      A\&A, 373, 827

   \bibitem[2000]{popescu} Popescu, C. C., Misiriotis, A., Kylafis, N. D.,
      et al. 2000,
      A\&A, 362, 138

   \bibitem[1995]{rauscher} Rauscher, B. J. 1995,
      AJ, 109, 1608

   \bibitem[1988]{rice} Rice, W., Lonsdale, C. J., Soifer, B. T., et al. 1988,
      ApJS, 68, 91

   \bibitem[1984]{rickard} Rickard, L. J., \& Harvey, P. M. 1984,
      AJ, 89, 1520

   \bibitem[1955]{salpeter} Salpeter, E. E. 1955,
      ApJ, 121, 161

   \bibitem[1998]{scalo} Scalo, J. M. 1998,
      in: The Stellar Initial Mass Function, eds.\ G. Gilmore, \& D. Howell
      (ASP Conf. Ser. 142)

   \bibitem[1997]{silk} Silk, J. 1997,
      ApJ, 481, 703

   \bibitem[1997]{smith} Smith, B. J., \& Madden, S. C. 1997,
      AJ, 114, 138 (SM97)

   \bibitem[1989]{soifer} Soifer, B. T., Boehmer, L., Neugebauer, G.,
      \& Sanders, D. B. 1989,
      AJ, 98, 766

   \bibitem[1985]{stacey1} Stacey, G. J. 1985,
      Ph.D. thesis, Cornell Univ.

   \bibitem[1991]{stacey2} Stacey, G. J., Geis, N., Genzel, R., et al. 1991,
      ApJ, 373, 423

   \bibitem[1982]{stauffer} Stauffer, J. R. 1982,
      ApJ, 262, 66

   \bibitem[1985]{tielens} Tielens, A. G. G. M., \& Hollenbach, D. 1985,
      ApJ, 291, 722

   \bibitem[2001]{tuffs} Tuffs, R. J., Popescu, C. C., Pierini, D.,
      et al. 2002,
      ApJS, 139, 37

   \bibitem[1995]{wolfire} Wolfire, M. G., Hollenbach, D., McKee, C. F.,
      et al. 1995,
      ApJ, 443, 152

   \bibitem[1994]{xu} Xu, C., Lisenfeld, U., V\"olk, H. J.,
      \& Wunderlich, E. 1994,
      A\&A, 282, 19 (X94)

\end{thebibliography}
\end{document}